\documentclass[10pt,a4paper,twocolumn,aps,prd,notitlepage,nofootinbib,showpacs,amsmath]{revtex4-1}
\usepackage{graphicx}

\usepackage{ulem}
\usepackage{color}
\newcommand{\pmin}[1]{{\color{red} #1}}

\begin{document}

\title{Two mass conjectures on axially symmetric black hole--disk systems.}
 
\author{Wojciech Kulczycki}
\author{Patryk Mach}
\author{Edward Malec}

\affiliation{Instytut Fizyki im.~Mariana Smoluchowskiego, Uniwersytet Jagiello\'{n}ski,
{\L }ojasiewicza 11, 30-348 Krak\'ow, Poland}
\begin{abstract}
 We analyze stationary self-gravitating disks around spinning black holes that satisfy the recently found 
 general-relativistic Keplerian rotation law.  There is a numerical evidence that  the  angular velocity,  circumferential radius and angular momenta yield a bound onto the asymptotic mass of the system.  This bound is proven analytically in the special case of massless disks of dust in  the Kerr spacetime.

\end{abstract}

\maketitle

\section{Introduction}

Angular velocities  $\Omega$ of   test bodies  in a nonrelativistic  Keplerian circular motion encode information about central  masses. The central mass is inferred to be $ \Omega^2 r_\mathrm{C}^{3}/G$, where $G $ is the gravitational coupling constant and $r_\mathrm{C}$ denotes the  circumferential radius of the orbit of the body. Recent numerical analysis suggests that one can estimate the total mass $m$ also for a system with selfgravitating fluids in Keplerian rotation around a spherical centre, $ \Omega^2 r_\mathrm{C}^{3}/G\le m$ \cite{MMP2012, MMP2013}. In this paper we shall investigate general-relativistic selfgravitating     disks  or toroids around spinning central black holes.   We shall propose two modified inequalities  for    selfgravitating rotating  systems in the general-relativistic Keplerian motion \cite{prd2018a,prd2018b}; they imply $ \Omega r_\mathrm{C}^{3/2} \le \sqrt{GM_\mathrm{ADM}} +\frac{2|a|}{\sqrt{r_\mathrm{C}}}$, where  $a$ and  $M_\mathrm{ADM}$ are the spin parameter  of the black hole and the asymptotic mass of    the whole configuration, respectively.  There is a rich numerical evidence, some reported in what follows, that supports our conjecture.

The question of estimating masses or determining rotation curves  in toroidal systems  has been addressed, within Newtonian physics, in the astrophysical literature \cite{Hure2002, Greenhill1996, Lodato2003,Hure2011, Hure2011a}. We would like to stress that our methods are different  and we work  within the general relativistic context.

 For simplicity, we assume only polytropic fluids  rotating around a (spinning or spinless) black hole, but we take into account the selfgravity of the fluid. Stationary axially symmetric systems of rotating barotropic  fluids need   the so-called rotation curve, that tells particles of fluid how to rotate. They are  described by the free-boundary elliptic systems of partial differential equations ---the shape of rotating fluids cannot be set \textit{a priori}, but comes with the solution. There exists a numerical technology   to deal with such problems, developed by many authors (a sample of sources: \cite{eriguchi,Hachisu,petroff,MSH,prd2016}), but the number of analytical results on the existence of solutions is small. There is a paper concerning self-gravitating fluids in Newtonian gravity \cite{Auchmuty} and a research on rigid rotations of self-gravitating gases in general relativity \cite{heilig,pfister1996,Schaudt}. 

In Newtonian gravity the Poincar\'{e}-Wavre theorem
  restricts    allowed  rotations curves  to 
$\Omega=\Omega(r_\mathrm{C})$, where the circumferential radius   $r_\mathrm{C}$ is equal to the distance from the rotation axis. They obviously include the Keplerian rotation law.  The  general-relativistic  rotation has been investigated since early 1970's --- primarily  a rigid rotation  \cite{Butterworth_Ipser,Bardeen_1970} and its modifications \cite{GYE,UTGHSTY,Uryu}.
 New general-relativistic differential rotation laws $j=j(\Omega)$, where $j$ denotes the specific angular momentum,  have recently been found  for stationary systems consisting of  self-gravitating toroids  around  spinless   \cite{MM,APP2015} and   spinning black holes \cite{prd2018a,prd2018b}.   They  describe, in particular,  the motion of  a  massless disks of dust, with   $\Omega = \sqrt{GM_\mathrm{ADM}}/r_\mathrm{C}^{3/2}$ in the case of the Schwarzschild geometry. In the nonrelativistic limit one has the monomial rotation law  $\Omega=w/r_\mathrm{C}^{\lambda}$
($0\le\lambda\le2$); that obviously includes  the Keplerian rotation law.  
 
 The order of the main part of this paper is as follows. Section II describes formalism. Section III contains  two main conjectures concerning masses of axially symmetric stationary systems with black holes. 
 In Section IV we show that test-like disks of dust satisfy the two mass conjectures in Schwarzschild and Kerr geometries. 
 The proof concerning the Kerr spacetime is relegated to the Appendix. Section V presents a   sample of numerical examples that confirm our two conjectures.
 
\section{Equations}

We assume a \textit{stationary} metric of the form 
\begin{align}
ds^{2} & =-\alpha^{2}{dt}^{2}+r^{2}\sin^{2}\theta\psi^{4}\left(d\varphi+\beta dt\right)^{2}\nonumber \\
 & \qquad+\psi^{4}e^{2q}\left(dr^{2}+r^{2}d\theta^{2}\right). \label{metric}
\end{align}
Here $t$ is the time coordinate, and $r$, $\theta$, $\varphi$
are spherical coordinates. In the general-relativistic part of  this paper the gravitational constant $G=1$ and the
speed of light $c=1$. We assume axial symmetry and employ the stress-momentum
tensor 
\[
T^{\alpha\beta}=\rho hu^{\alpha}u^{\beta}+pg^{\alpha\beta},
\]
where $\rho$ is the baryonic rest-mass density, $h$ is the specific
enthalpy, and $p$ is the pressure. Metric functions $\alpha $,
$\psi $, $q $ and $\beta $ in \eqref{metric}
depend on $r$ and $\theta$ only.

The following method can be applied to any barotropic equation of
state\pmin{,} but we will deal with polytropes $p(\rho)=K\rho^{\gamma}$.
Then one has the specific enthalpy 
\[
h(\rho)=1+\frac{\gamma p}{(\gamma-1)\rho}.
\]
The 4-velocity $(u^{\alpha})=(u^{t},0,0,u^{\varphi})$ is normalized,
$g_{\alpha\beta}u^{\alpha}u^{\beta}=-1$. The coordinate angular velocity
reads 
\begin{equation}
\Omega=\frac{u^{\varphi}}{u^{t}}.\label{Omega_def}
\end{equation}
  We  define the angular momentum per unit
inertial mass $\rho h$ \cite{FM} 
\begin{equation}
j\equiv u_{\varphi}u^{t}.\label{j_def}
\end{equation}
It is  known since early 1970's that general-relativistic Euler equations are solvable under the condition that $j\equiv j(\Omega)$ \cite{Butterworth_Ipser,Bardeen_1970}.
Within the fluid region, the Euler equations $\nabla_{\mu}T^{\mu\nu}=0$
can be integrated,
\begin{equation}
\int j(\Omega) d\Omega+\ln\left(\frac{h}{u^{t}}\right)=C.\label{uf}
\end{equation}

In \cite{MM} we have had the rotation law 
\begin{equation}
j(\Omega)\equiv\frac{\tilde{w}^{1-\delta}\Omega^{\delta}}{1-\kappa\tilde{w}^{1-\delta}\Omega^{1+\delta}+\Psi};\label{prime}
\end{equation}
here $\Psi$ is of the order of the binding energy per unit baryonic
mass and  $w$, $\delta$, and $\kappa=(1-3\delta)/(1+\delta)$ are parameters. This law was obtained in a procedure involving  an ``educated guess-work''   \cite{MM}.

 Equation (\ref{prime}) can be transformed (through a rescaling of $\tilde w$) into 
   \[
j(\Omega)\equiv\frac{w^{1-\delta}\Omega^{\delta}}{1-\kappa w^{1-\delta}\Omega^{1+\delta}}=\left(-\kappa\,\Omega+w^{\delta-1}\Omega^{-\delta}\right)^{-1},
\]
The Keplerian rotation corresponds to the parameter 
$\delta =-1/3$ and $\kappa =3$; it is interpreted as  a rotation law  for the fluid around a spin-less black hole.

The rotation law describing motion of gaseous tori around spinning black holes,
\begin{equation}
j\left(\Omega\right)=-\frac{1}{2}\frac{d}{d\Omega}\ln\left\{ 1- \left[\tilde a^{2}\Omega^{2}+3w^{\frac{4}{3}}\Omega^{\frac{2}{3}}\left(1-\tilde a\Omega\right)^{\frac{4}{3}}\right]\right\} ,
\label{j(O)-1}
\end{equation}
 has been found in \cite{prd2018a, prd2018b}.   Here $\tilde a$ is a kind of a bare spin parameter of a black hole (see a comment below); it coincides with the Kerr spin parameter for massless disks of dust.
  
The rotation curves---angular velocities as functions of spatial coordinates $\Omega(r,\theta )$---can  be recovered from   Eq.\ (\ref{j_def}),
\begin{equation}
j(\Omega)=\frac{V^{2}}{\left(\Omega+\beta\right)\left(1-V^{2}\right)}.\label{rotation_law}
\end{equation}
Here  the squared  linear velocity is given by
\[
V^{2}=r^{2}\sin^{2}\theta\left(\Omega+\beta\right)^{2}\frac{\psi^{4}}{\alpha^{2}}.
\]

 Following \cite{MSH} we introduce the central black hole using the puncture method \cite{BrandtSeidel1995}. The   black hole is surrounded by a minimal two-surface $S_{\mathrm{BH}}$ (the horizon)  embedded in a fixed hypersurface of constant time, and
located at  $r=r_{\mathrm{s}} = \sqrt{m^2 - \tilde a^2}/2$, where $m$ is a mass parameter. Its area defines the irreducible
mass $M_{\mathrm{irr}}=\sqrt{\frac{A_{\mathrm{H}}}{16\pi}}$ and its
angular momentum $J_{\mathrm{BH}}$ follows from the Komar expression
\begin{equation}
J_{\mathrm{BH}}=\frac{1}{4}\int_{0}^{\pi/2}\frac{r^{4}\psi^{6}}{\alpha}\partial_{r}\beta\sin^{3}\theta d\theta.
\end{equation}
In this construction  the angular momentum is given rigidly on the event horizon $S_{\mathrm{BH}}$, in terms of 
   data taken from the Kerr solution  and
independently of the content of mass in a torus,  $J_\mathrm{BH} = m \tilde a$ \cite{MSH}. The mass of the black
hole is then defined in terms of the irreducible mass and the angular momentum,
\begin{equation}
M_{\mathrm{BH}}=M_{\mathrm{irr}}\sqrt{1+\frac{J_{\mathrm{BH}}^{2}}{4M_{\mathrm{irr}}^{4}}}.
\end{equation}
 We define the black hole spin parameter as $a = J_\mathrm{BH}/M_\mathrm{BH}$. If the disk is sufficiently massive (self-gravitating), we have in general $a \neq \tilde a$  \cite{prd2018a, prd2018b}. If the self-gravity of the torus can be neglected, then $M_\mathrm{BH} = m$, $a = \tilde a$, and the metric of the spacetime coincides (by construction) with the Kerr solution.

 Asymptotic (total) mass $M_\mathrm{ADM}$  and angular momentum $J_\mathrm{ADM}$ can be defined as apropriate Arnowitt-Deser-Misner charges, and they can be computed by means of corresponding volume integrals \cite{MSH}.

A circumferential radius corresponding to the circle $r=\mathrm{const}$ on the symmetry plane $\theta =\pi /2$
is given by
\begin{equation}
r_\mathrm{C}=r\psi^2.
\label{rad}
\end{equation}
    
 The numerical part of this paper  is based on the scheme described in \cite{prd2018a,prd2018b}.
In the rest of the paper we always assume that $\Omega>0$. Corotating
disks have $a>0$, while counterrotating disks have negative spins:
$a<0$.

\section{Two mass conjectures } 
 
 General-relativistic Keplerian systems with tori are characterized by their asymptotic masses  $M_\mathrm{ADM}$ and angular momenta  $J_\mathrm{ADM}$,
 and the quasilocal characteristics of central black holes---the mass $M_\mathrm{BH}$ and the angular momentum $J_\mathrm{BH}$.  It is clear that in the Newtonian limit we should get  $ \Omega r_\mathrm{C}^{3/2}\le \sqrt{m}$ \cite{MMP2012, MMP2013}.  From this, and from the dimensional analysis, one may guess that $  \Omega r_\mathrm{C}^{3/2}\le \sqrt{M_\mathrm{ADM}}+2\frac{|J_\mathrm{X }|}{m_\mathrm{Y}\sqrt{r_\mathrm{C}}}  $,  where $\mathrm{X}, \mathrm{Y}= \mathrm{BH}, \mathrm{ADM}$.   
The asymptotic mass is larger than the mass of the central black hole, but the angular momentum is not monotonic --- the asymptotic angular momentum might be smaller than $J_\mathrm{BH}$. That  means that  there are two independent  possibilities.

\textbf{Conjecture 1.}  

\begin{equation}
\Omega r_\mathrm{C}^{3/2}\le \sqrt{M_\mathrm{ADM}}+\frac{2|J_\mathrm{ADM}|}{M_\mathrm{ADM}\sqrt{r_\mathrm{C}}}. 
\label{c1}
\end{equation}

\textbf{Conjecture 2.}
  
\begin{equation}
\Omega r_\mathrm{C}^{3/2}\le \sqrt{M_\mathrm{ADM}}+\frac{2\left|J_{\mathrm{BH}}\right|}{M_{\mathrm{ADM}}\sqrt{r_{\mathrm{C}}}}.
\label{c2}
\end{equation}

Notice that  $\frac{\left|J_{\mathrm{BH}}\right|}{M_{\mathrm{ADM}} }\le  |a| $, where $a$ is the black hole spin parameter.
Thus Conjecture 2 implies $ \Omega r_\mathrm{C}^{3/2} \le \sqrt{M_\mathrm{ADM}} +\frac{2|a|}{\sqrt{r_\mathrm{C}}}$.

Numerical calculations reported in Sec. VI  suggest the validity of both Conjectures.
They   coincide for test disks of dust in the Kerr geometry, since then $M_\mathrm{BH}=
 M_\mathrm{ADM}$ and  $J_\mathrm{ADM}= J_\mathrm{BH}$. The   inequality reduces to the equality in the Schwarzschild spacetime 
 and it is satisfied   in Kerr spacetimes (see the Appendix for an algebraic  proof). 
 
 The inspection of  both  inequalities (\ref{c1}) and (\ref{c2}) shows that far from the center the  corresponding expressions on the right-hand sides are expected to be constant; if a toroid is light and large, then $\Omega r_\mathrm{C}^{3/2}$ is expected to be close to $  \sqrt{M_\mathrm{ADM}}$ (see Sec. V). The inspection of these inequalities in the interior  might give some information about angular momentum.
   
\section{Angular momentum and mass estimates in Kerr and Schwarzschild spacetimes}

 \subsection{Kerr geometry in conformal coordinates}

Define 
\begin{equation}
 r_\mathrm{K} = r \left( 1 + \frac{m}{r} + \frac{m^2 - a^2}{4 r^2} \right), 
 \label{rk}
 \end{equation}
 
\[ \Delta_\mathrm{K} =r_\mathrm{K}^2 -2r_\mathrm{K}+a^2,\]
and
\[ \Sigma_\mathrm{K} = r_\mathrm{K}^2 + a^2 \cos^2 \theta. \]

 The Kerr metric can be written in form (\ref{metric}) as follows \cite{BrandtSeidel1995, MSH}. The conformal factor $\psi_\mathrm{K}  $ reads 
 \begin{equation}
 \psi_\mathrm{K}  = \frac{1}{\sqrt{r}}\Bigl( r^2_\mathrm{K}  +a^2 +2ma^2\frac{r_\mathrm{K}\sin^2\theta  }{\Sigma_\mathrm{K}}\Bigr)^{1/4}.
 \label{psik}
 \end{equation}
The only component $\beta_\mathrm{K} $ of the shift vector  is given by 
 \begin{equation}
 \beta_\mathrm{K}  =-\frac{2mar_\mathrm{K}}{(r^2_\mathrm{K}+a^2)\Sigma_\mathrm{K} +2ma^2r_\mathrm{K} \sin^2\theta}.
 \label{betak}
 \end{equation}
 Finally, the functions $\alpha_\mathrm{K} $ and $ q_\mathrm{K}$ are defined as
 \begin{eqnarray}
 \alpha_\mathrm{K}  &=& \Bigl( \frac{ \Sigma_\mathrm{K} \Delta_\mathrm{K}}{(r_\mathrm{K}^2+a^2)\Sigma_\mathrm{K}+2ma^2r_\mathrm{K}}\sin^2\theta\Bigr)^{1/2},
 \nonumber \\
 e^{q_\mathrm{K}} &=&\frac{\Sigma_\mathrm{K}}{\sqrt{(r^2_\mathrm{K}+a^2)\Sigma_\mathrm{K} +2ma^2r_\mathrm{K} \sin^2\theta}}.
 \label{alphaq}
 \end{eqnarray}

The surface $r = r_\mathrm{s}\equiv\frac{1}{2}\sqrt{m^2-a^2}$ is an apparent horizon, that coincides with the event horizon.

  Test particles can rotate along circular orbits $r=\mathrm{const}$ in the Kerr geometry. That implies the existence of a test-like disk made of dust, that moves circularly and lies in the plane $\theta =\pi /2$.
 Its angular velocity reads
 \begin{equation}
 \Omega(r)=\frac{8r^{3/2}}{\left((2r+1)^{2}-a^{2}\right)^{3/2}+8ar^{3/2}}.
\label{omega}
\end{equation}
The dragging angular velocity of the Kerr space-time is given by $\Omega_\mathrm{d}=-\beta_\mathrm{K} $, that is
 \begin{equation}  
 \Omega_\mathrm{d} =\frac{2ma }{r_\mathrm{K}(r^2_\mathrm{K}+a^2)+2ma^2}.
\label{omegad}
\end{equation}
  The circumferential radius of this circular orbit is equal to
\begin{equation}
r_\mathrm{C}=r\psi^2=\sqrt{r^2_\mathrm{K}+a^2+\frac{2ma^2}{r_\mathrm{K}}}
\label{radc}
\end{equation}
It is clear that 
\begin{equation}  
 \Omega_\mathrm{d} =\frac{2ma }{r_\mathrm{K}r^2_\mathrm{C} }.      
\label{omegadd}
\end{equation}

It appears that the product $\Omega r_\mathrm{C}^{3/2}$ exceeds the value  $\sqrt{m}$ for all strictly negative spin parameters $a$ and for   $a>0.9525$. 
There exists, however, a modified inequality that takes into account the dragging effects and that is always true: \begin{equation}
\Omega r_\mathrm{C}^{3/2}\le \sqrt{m}+  |\Omega_\mathrm{d}|
r_\mathrm{C}^{3/2} =\sqrt{m}+\frac{2m|a| }{r_\mathrm{K}r^{1/2}_\mathrm{C}} .  
\label{inood}
\end{equation}
The proof of (\ref{inood}) is relegated to the Appendix.

The quantity $m|a|$ is  the absolute value $|J_\mathrm{BH}|$ of the angular momentum, while $r_\mathrm{K}\ge m$ outside the region 
encircled by the trapped surface  $r=r_\mathrm{s}  $. Thus for a disk in Keplerian motion around a Kerr black hole we have 
\begin{equation}
 \Omega r_\mathrm{C}^{3/2} \le \sqrt{m}+2\frac{J_\mathrm{BH}}{m\sqrt{r_\mathrm{C}}} = \sqrt{m}+2\frac{|a|}{\sqrt{r_\mathrm{C}}}.
 \label{ik2}
 \end{equation}
This agrees with  both inequalities (\ref{c1}) and (\ref{c2}),
since they coincide in this case.

\subsection{Schwarzschild geometry} 

Schwarzschild space-time in conformal coordinates is given by metric functions of the preceding subsection assuming  the spin parameter $a=0$.
Test bodies in a Schwarzschild space-time can move on circular orbits with the angular velocity $\Omega = \frac{\sqrt{m}}{r_\mathrm{C}^{3/2}}  $. Thus there exist massless
 disks, consisting of particles of dust,  with the rotation law $\Omega = \frac{\sqrt{m_{}}}{r_\mathrm{C}^{3/2}}  $. This implies   the strict equality $\Omega r_\mathrm{C}^{3/2}= \sqrt{m}$; the   inequalities (\ref{c1}) and (\ref{c2}) coincide, and they are saturated.

\section{Numerical results}

 In this section we deal with numerical solutions describing self-gravitating fluids around black holes. The numerical method was described in detail in \cite{prd2018a,prd2018b}. 

 It is convenient to rewrite inequalities (\ref{c1}) and (\ref{c2}) in the following form.
\begin{enumerate}
\item Conjecture 1.
\begin{equation}
\max\Bigl((\Omega-  \Omega_{\mathrm{d},1}
 )r_{\mathrm{C}}^{3/2}\Bigr)\leq\sqrt{M_{\mathrm{ADM}}};\label{ineq1}
\end{equation}
  here
$\Omega_{\mathrm{d},1}=\frac{2\left|J_\mathrm{ADM}\right|}{M_{\mathrm{ADM}}r_{\mathrm{C}}^{2}}$.
\item Conjecture 2.
\begin{equation}
\max\Bigl((\Omega-  \Omega_{\mathrm{d},2}
 )r_{\mathrm{C}}^{3/2}\Bigr)\leq\sqrt{M_{\mathrm{ADM}}};\label{ineq2}
\end{equation}
here $\Omega_{\mathrm{d},2}= \frac{2\left|J_{\mathrm{BH}}\right|}{M_{\mathrm{ADM}}r_{\mathrm{C}}^{2}}$.
\end{enumerate}
 
We performed  a large number of numerical calculations, with equations of state  $p=K\rho^{4/3}$ and $p=K\rho^{5/3}$,
for a range of  values of the spin parameter  $a$, the radius of the inner boundary $r_{1}$ and for several values of the maximal mass density $\rho_{\mathrm{max}}$. In this paper we report only results concerning the polytrope  with the polytropic index $\gamma =4/3$, but the other case gives similar results.
In    numerical solutions of  Figs. 1 and 2  we  fix the coordinate radius of the  disk's outer boundary $r_2=20$,
but the inner radius $r_1$  is changed, within limits shown in captions of   Figures. Figures 3--5 are dedicated to the analysis of  $\Omega r_\mathrm{C}^{3/2}$
in specific solutions.   In all cases we calculated values of the left hand sides of (\ref{ineq1}) and (\ref{ineq2}) in the symmetry plane  $\theta = \pi/2$.

      Figure 1 summarizes results of more than four hundred (counter-rotating) numerical solutions. The diagram shows maximal values of the left hand sides of (\ref{ineq1}) and (\ref{ineq2}) in units of the square root of the asymptotic mass.   There is a  sharp spike in the diagram  corresponding to Conjecture 1 (see (\ref{ineq1})); around $r_1=11.7$ the angular momentum of the disk   cancels the black hole spin.  The asymptotic angular momentum    $J_\mathrm{ADM}$   decreases, vanishes at the top of the spike     and becomes more and more negative.    Conjecture 2 is also satisfied, but the expression $\mathrm{max} ((\Omega - \Omega_{\mathrm{d},2})r_\mathrm{C}^{3/2})/\sqrt{M_\mathrm{ADM}}$    changes only moderately.    Figure  2   presents results of more than seven hundred (co-rotating) numerical solutions.        It is clear that both proposed conjectures, 1 and 2, are valid in our numerical calculations, for co- and counter- rotating systems. 
 \begin{figure}[ht]
\includegraphics[width=1\columnwidth]{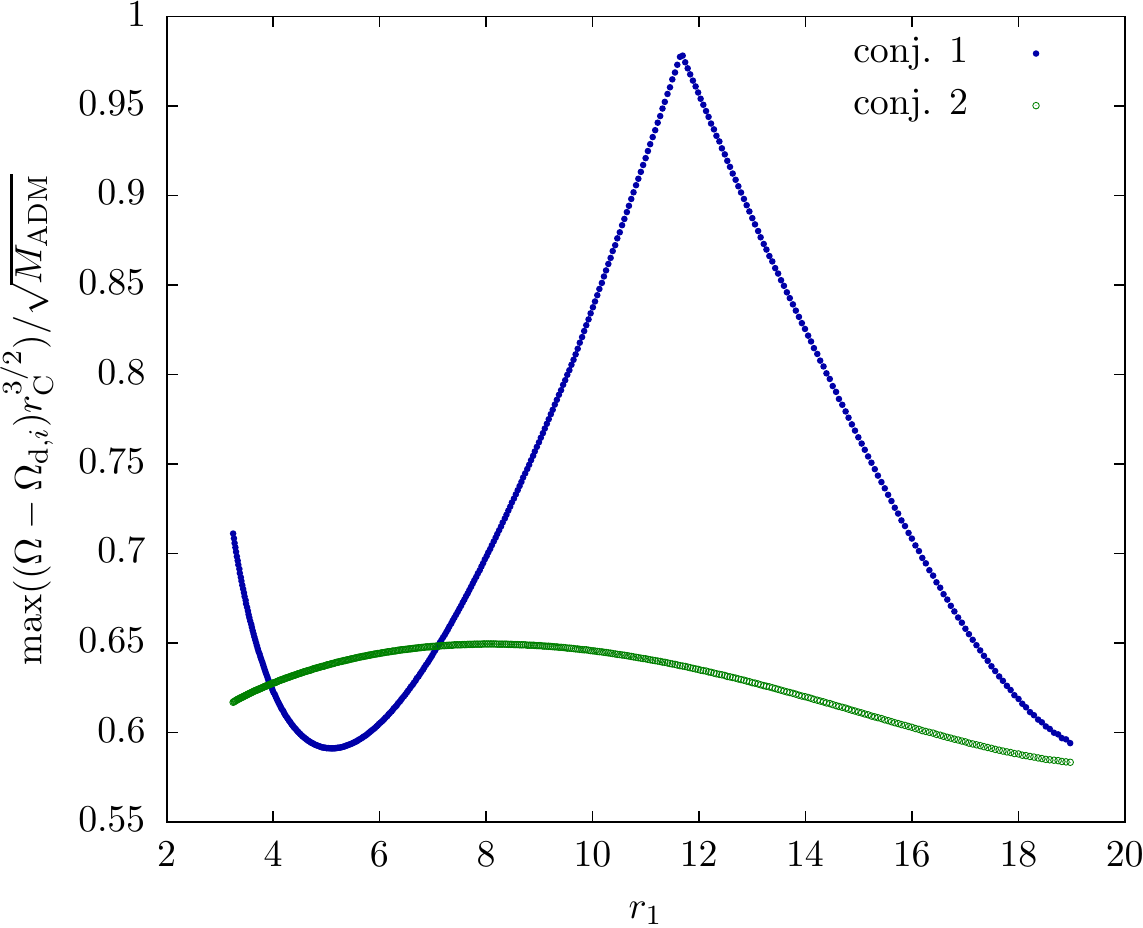}
\caption{Maximal values of left   hand-sides of Conjectures 1 (conj. 1) and 2 (conj. 2)   for $\gamma=4/3$, $\tilde a=-0.99$ and $\rho_{\mathrm{max}}=3.0\times10^{-4}$. Here $r_2=20$
and   $r_1$ varies from 6.01 to 18.97. Asymptotic masses are in the range $(1.005, 1.46)$. Each point on the diagram corresponds to a solution.}
\end{figure}

\begin{figure}[ht]
\includegraphics[width=1\columnwidth]{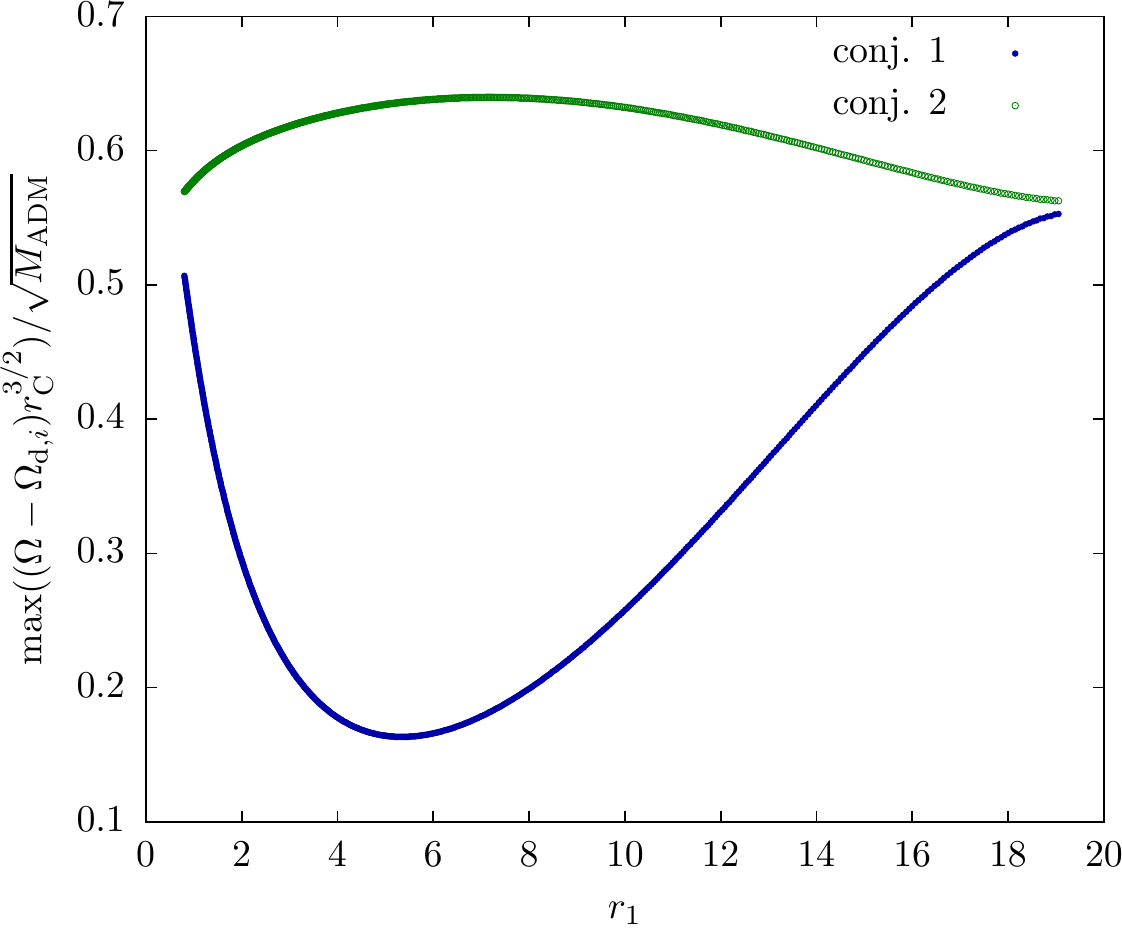}
\caption{     Maximal values of left   hand-sides of Conjectures 1 and 2   for $\gamma=4/3$, $\tilde a=0.99$ and $\rho_{\mathrm{max}}=3.0\times10^{-4}$. Here $r_2=20$
and   $r_1$ varies from 0.805 to  19.05. Asymptotic  masses are in the range $( 1.005,1.46 )$.    Each point on the diagram corresponds to a solution. }
\end{figure}     
      
  Figure 3  shows the behaviour of the product 
  $\Omega r_\mathrm{C}^{3/2}$ on the symmetry plane of a    compact disk---its outer circumferential  radius is less than 21.6 and the asymptotic mass $M_\mathrm{ADM}\in (1.45,1.47)$ while  the mass of the central  black hole is $M_\mathrm{BH}=1.02$. We display two curves, for $a=0.49$ and $a=-0.485$. The dependence on the spin is more pronounced in disk's interior and becomes negligible in disk's peripherals.

\begin{figure}[ht]
\includegraphics[width=1\columnwidth]{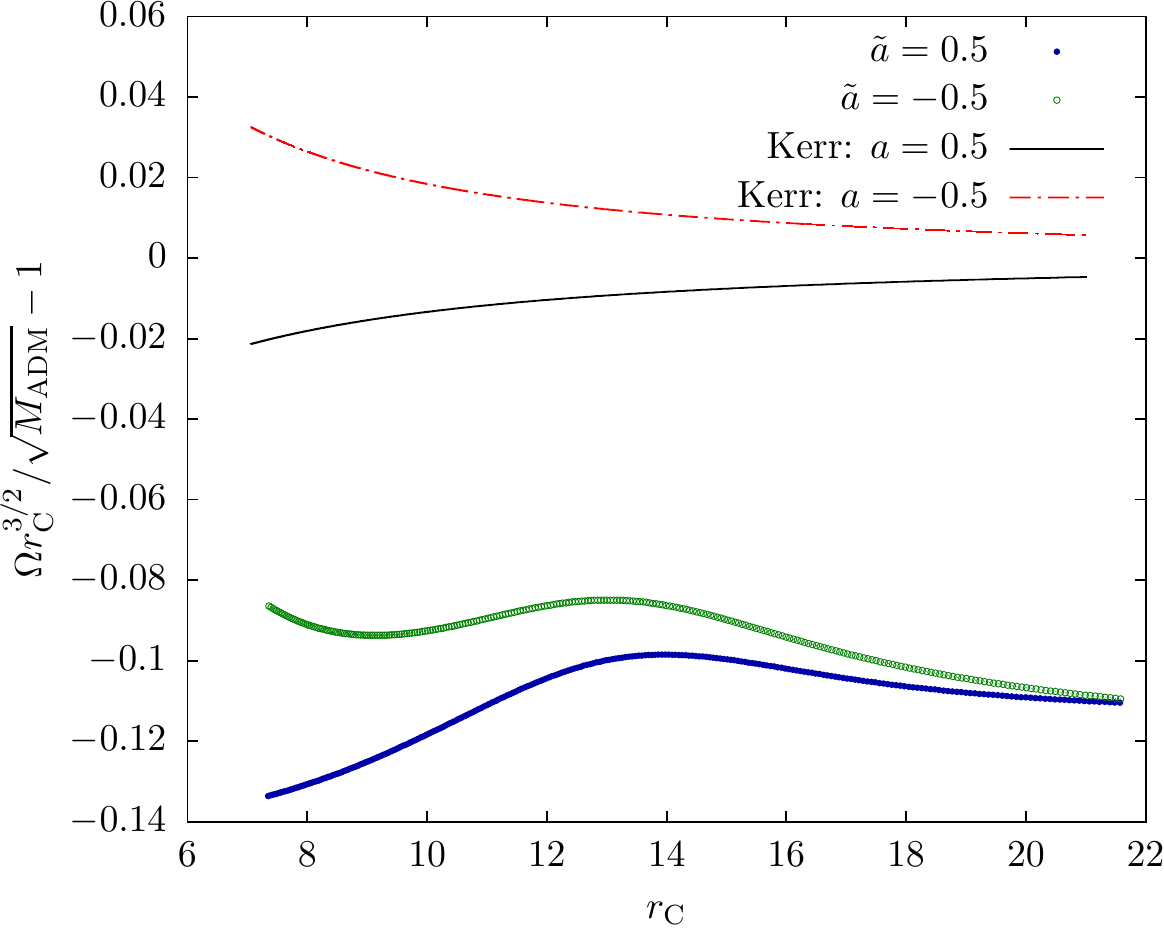}
\caption{    The plot of   $\frac{\Omega r_\mathrm{C}^{3/2}}{M_\mathrm{ADM}}-1$   as function of $r_\mathrm{C}$.
Here      $\gamma=4/3$
and $\rho_{\mathrm{max}}=3.0\times10^{-4}$.
i)  Blue line:  $a=0.49$, $r_\mathrm{C}=\in (7.35,  21.57)$, $M_\mathrm{ADM}=1.45$ and $M_\mathrm{BH}=1.02$. 
ii) Green line:  $a= -0.485$,  $r_\mathrm{C}\in (7.36,  21.58)$, $M_\mathrm{ADM}=1.47$ $M_\mathrm{BH}=1.03$. 
iii) For  the comparison, the plot of $\frac{\Omega r_\mathrm{C}^{3/2}}{m }-1$ for a massless disk in the Kerr geometry, with $m=1$. }
\end{figure}

\begin{figure}[ht]
\includegraphics[width=1\columnwidth]{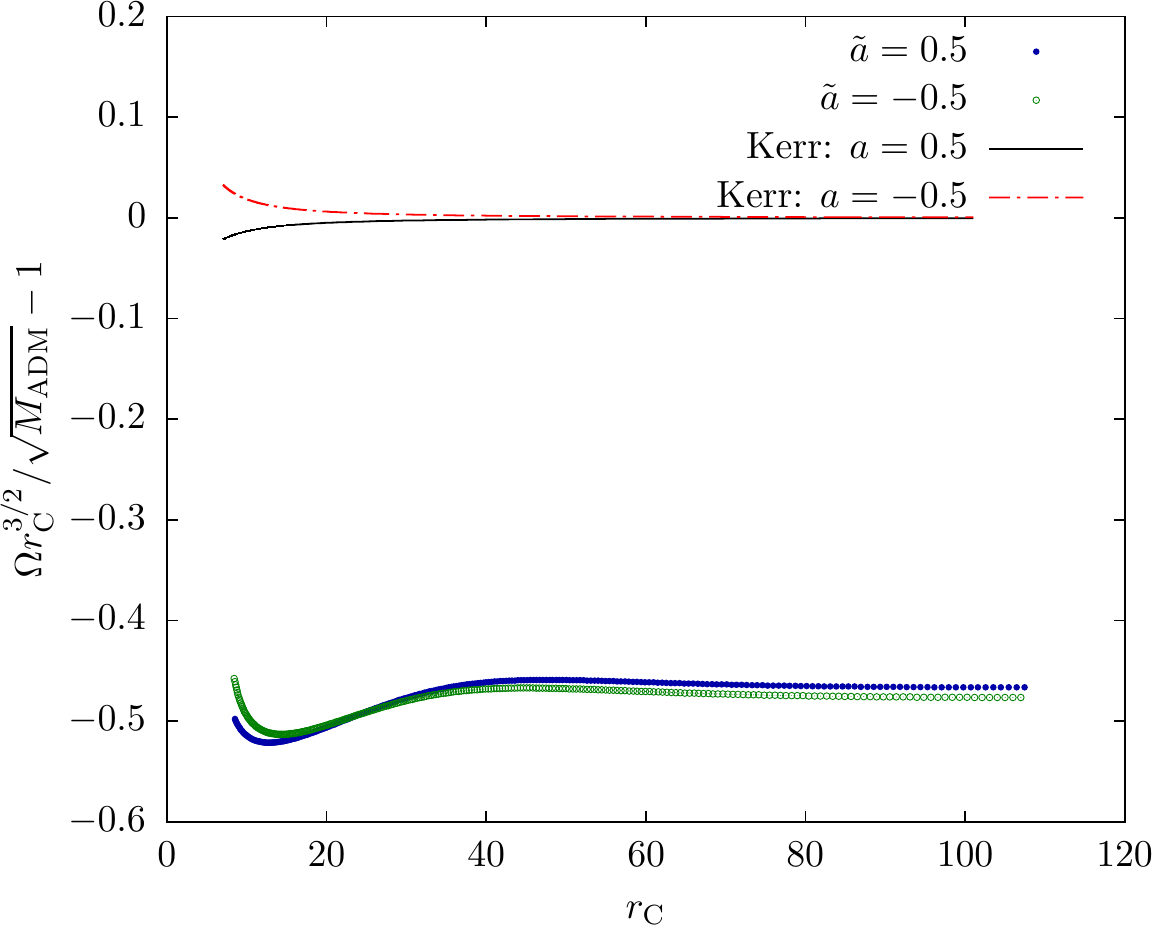}
\caption{   The plot of   $\frac{\Omega r_\mathrm{C}^{3/2}}{M_\mathrm{ADM}}-1$   as function of $r_\mathrm{C}$.
Here  $\gamma=4/3$, $r_\mathrm{C}\in (7.27,  102.32)$,  $M_\mathrm{ADM}=2$ and $M_\mathrm{BH}=1.02$.
i)  Blue line:  $a=0.49$, $\rho_{\mathrm{max}}=7.47\times10^{-6}$.
ii) Green line:  $a= -0.49$, $\rho_{\mathrm{max}}=7.74\times10^{-6}$.
iii) For  the comparison, the plot of $\frac{\Omega r_\mathrm{C}^{3/2}}{m }-1$ for a massless disk in the Kerr geometry, with $m=1$.  }
\end{figure}

 Figure 4  shows the behaviour of the product 
  $\Omega r_\mathrm{C}^{3/2}$ on the symmetry plane of the disk. The disk    is quite extended and somewhat heavier ---its outer circumferential  radius is larger than 100 and the asymptotic mass $M_\mathrm{ADM}=2$ while  the mass of the black hole is $M_\mathrm{BH}=1.02$. We display two curves, for $a=0.49$ and $a=-0.49$. The dependence on the spin is seen   in disk's interior and becomes negligible in disk's peripherals.    
  
\begin{figure}[ht]
\includegraphics[width=1\columnwidth]{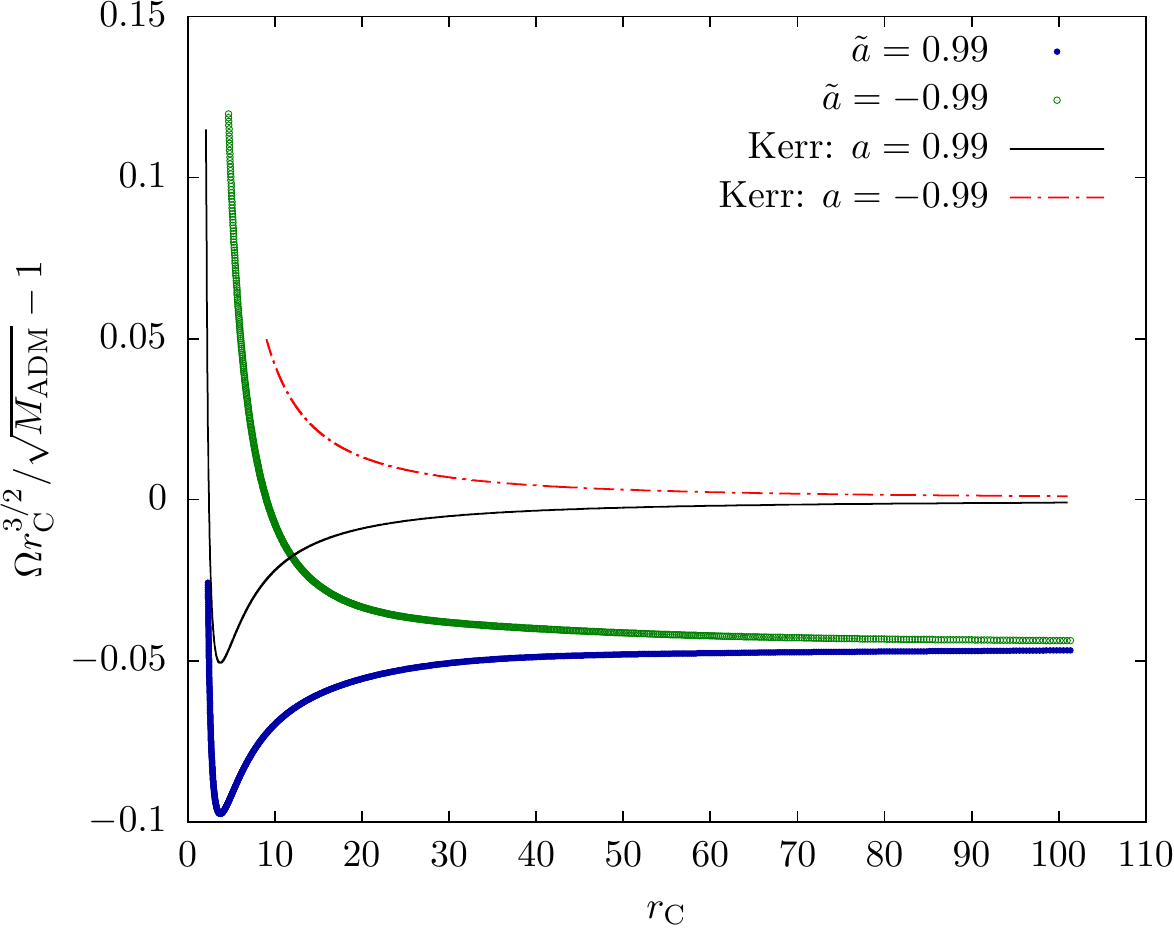}
\caption{   The plot of   $\frac{\Omega r_\mathrm{C}^{3/2}}{M_\mathrm{ADM}}-1$   as function of $r_\mathrm{C}$.
Here  $\gamma=4/3$,   $M_\mathrm{ADM}=1.1$ and $M_\mathrm{BH}=1.00$.
i)  Green line:  $a=-0.99$, $\rho_{\mathrm{max}}=1.92\times10^{-6}$, $r_\mathrm{C}\in (4.68,101.36)$.
ii) Blue  line:  $a= 0.99$, $\rho_{\mathrm{max}}=2.5\times10^{-6}$, $r_\mathrm{C}\in (2.32,101.36)$.
iii) For  the comparison, the plot of $\frac{\Omega r_\mathrm{C}^{3/2}}{m }-1$ for a massless disk in the Kerr geometry ($m=1$).  For $a=-0.99$: $r_\mathrm{C}\in  (9.04,101.00)  $, while for $a=0.99$: $r_\mathrm{C}\in  (2.108,101.00)  $. 
 }
\end{figure}

  Figure 5 shows the same as Fig. 4, but the spin is higher.
 Again the disk    is   extended  but it is light---its outer circumferential  radius is larger than 100 and the asymptotic mass $M_\mathrm{ADM}=1.1$   while  the mass of the black hole is $M_\mathrm{BH}=1.00$. We display two curves, for $a=0.99$ and $a=-0.99$. A strong  dependence on the spin is seen   in disk's interior, but it is noticeable even in disk's peripherals.   Notice that $\sup (\Omega r_\mathrm{C}^{3/2})>1.1$ for the two counter-rotating branches, but only the massless co-rotating branch has $\Omega r_\mathrm{C}^{3/2} $ exceeding  1.

It is instructive  to  compare the two self-gravitating solutions  in Fig. 3 with the two massless disks of dust in the Kerr geometry. It is clear that the self-gravity  merely  pushes down the value of   $\Omega r_\mathrm{C}^{3/2}/M_\mathrm{ADM}-1$  obtained  for dust solutions in Kerr space-times. The branch corresponding to the counter-rotation remains above the co-rotating one, but they almost coincide at the outer boundary.
A similar picture is seen in Fig. 5.
We see the same effect    in  Fig. 4, but with a more pronounced shift downwards. The co-rotating and counter-rotating branches cross around $r_\mathrm{C}\approx 30$ and then their positions do reverse.       Thus the more massive is the rotating toroid, the stronger the self-gravity impacts the quantity $\Omega r_\mathrm{C}^{3/2}$. Figure 5 demonstrates yet another influence of    the self-gravity;  the inner  boundary of the co-rotating toroid shifts upwards from $r_\mathrm{C}=2.108$ (in the Kerr spacetime)  to $r_\mathrm{C}=2.32$ and of the counter-rotating toroid moves downwards from   $r_\mathrm{C}=9.04$ (in the  Kerr spacetime)  to $r_\mathrm{C}=4.68$.

\section{Summary}

The two conjectures on masses of rotating systems are based  partly  on analytic arguments and partly on extensive numerical data. Their proof would pose a serious analytic challenge. 

We envisage two further applications. The first concerns astrophysics. Rotating axially symmetric systems  are quite common. They are  known to exist in some active galactic nuclei.  Particularly interesting are  those containing supermasers. Our inequalities might be useful in extracting information about masses and angular momenta, provided that more information on modelling of AGN's with toroids becomes available.

The two statements of  (\ref{c1}) and (\ref{c2}) can be useful in estimating the amount of angular momentum within a fixed 
volume. There is already a formidable work done in this direction \cite{Dain, Xie_Khuri}, although results are far from being precise \cite{prd2016}. We think that (\ref{c1}) and (\ref{c2}) are true and they would lead to substantial improvements of 
present estimates.

\section{Appendix. Proof of the mass inequality in the Kerr geometry}

We shall do this proof in Boyer-Lindquist coordinates $(r_\mathrm{K}, \theta, \phi )$, where the Boyer-Lindquist radius $r_\mathrm{K}$ relates to the conformal radius $r$ via formula (\ref{rk}) \cite{BrandtSeidel1995}. The metric takes on the standard form

\begin{eqnarray}
ds^{2}&=&-\left(1-\frac{2mr_{\textrm{K}}}{\Sigma_{\textrm{K}}}\right)dt^{2}+\frac{\Sigma_{\textrm{K}}}{\Delta_{\textrm{K}}}dr_{\textrm{K}}^{2}+\nonumber \\
&&\Sigma_{\textrm{K}} d\theta^{2}+\frac{\Delta_{\textrm{K}}\Sigma_{\textrm{K}}+2mr_{\textrm{K}}(r_{\textrm{K}}^{2}+a^{2})}{\Sigma_{\textrm{K}}}\sin^{2}\theta d\phi^{2}-\nonumber\\
&&\frac{4mr_{\textrm{K}}a\sin^{2}\theta}{\Sigma_{\textrm{K}}}dtd\phi,\label{metryka_Kerra_BL}
\end{eqnarray}

\noindent where $\Sigma_{\textrm{K}}=r_{\textrm{K}}^{2}+a^{2}\cos^{2}\theta$, $\Delta_{\textrm{K}}=r_{\textrm{K}}^{2}-2mr_{\textrm{K}}+a^{2}$.

 The inequalities of Section V are expressed in terms of
geometric quantities, hence they are independent of the choice of coordinates within a fixed   Cauchy slice.
We shall  consider a massless disk, that is  located in the symmetry plane $\theta =\pi /2$.

\subsection{Geometric quantities}

The angular velocity $\Omega$ reads
\begin{equation}
\Omega=\frac{u^{\phi}}{u^{t}}=\frac{\sqrt{m}}{r_{\textrm{K}}^{3/2}+a\sqrt{m}}.\label{dowod_nier_Kerr_Omega}
\end{equation}
The circular radius  $r_\mathrm{C}$  can be expressed as follows, in the symmetry plane:
\begin{equation}
r_{\textrm{C}} =\sqrt{r_{\textrm{K}}^{2}+a^{2}+\frac{2ma^{2}}{r_{\textrm{K}}}}.\label{dowod_nier_Kerr_rc}
\end{equation}
The  angular velocity due to dragging plays a significant role in the calculation. It is given by 
\begin{equation}
\Omega_{\textrm{d}}=\frac{2ma}{r_{\textrm{C}}^{2}r_{\textrm{K}}}.\label{dowod_nier_Kerr_Omegaw}
\end{equation}
The irreducible mass of the Kerr black hole reads
\begin{equation}
M_{\textrm{irr}}=\frac{m}{2}\sqrt{2\left(1+\sqrt{1-\frac{a^{2}}{m^{2}}}\right)}.\label{dowod_nier_Kerr_Mirr}
\end{equation}
The quantity $r_{\textrm{ISCO}}$ denotes the coordinate radius of the innermost stable circular orbit (ISCO),
that depends on $a$ and $m$. In the case of co-rotation  $r_{\textrm{ISCO}}\ge m$ (with the equality when 
$a=1$) while in the case of counter-rotation   $6m\le r_{\textrm{ISCO}}\le 9m$ (with the upper bound saturated for $a=-1$).

\subsection{Proofs}

We shall prove analytically the validity of the following inequality, provided that the areal radius is not smaller than  
$r_{\textrm{ISCO}}$: $r_{\textrm{K}}\geq r_{\textrm{ISCO}}$:
\begin{equation}
\Omega r_{\textrm{C}}^{3/2}  \leq \sqrt{m}+|\Omega_{\textrm{d}}|r_{\textrm{C}}^{3/2}.
\label{nierownosc_Kerrb}
\end{equation}
For the simplicity, but without the loss of generality, we shall put  $m=1$.
 
We divide both sides of $(\ref{nierownosc_Kerrb})$
by $r_{\textrm{C}}^{3/2}$; that leads to
\begin{equation}
\Omega-|\Omega_{\textrm{d}}|\leq\sqrt{\frac{1}{r_{\textrm{C}}^{3}}}.\label{nier_Kerr_gor}
\end{equation}
Using  (\ref{rk}, \ref{dowod_nier_Kerr_Omega}--\ref{dowod_nier_Kerr_Mirr}), one arrives at 
\begin{eqnarray}
\lefteqn{\frac{1}{r_{\textrm{K}}^{3/2}+a}-\frac{2|a|}{r_{\textrm{K}}\left(r_{\textrm{K}}^{2}+a^{2}\right)+2a^{2}}\leq} \nonumber \\
&& \leq\left(a^{2}+\frac{2a^{2}}{r_{\textrm{K}}}+r_{\textrm{K}}^{2}\right)^{-3/4}.\label{nier_Kerr_gor_wstawione}
\end{eqnarray}

\paragraph{The case $a=0$} corresponds to the Schwarzschild geometry and it has been already discussed.
\paragraph{In the corotating case, $a>0$,}   inequality $(\ref{nier_Kerr_gor_wstawione})$ takes the form
\begin{eqnarray}
\lefteqn{\frac{1}{r_{\textrm{K}}^{3/2}+a}-\frac{2a}{r_{\textrm{K}}\left(r_{\textrm{K}}^{2}+a^{2}\right)+2a^{2}}\leq} \nonumber \\
&& \leq\left(a^{2}+\frac{2a^{2}}{r_{\textrm{K}}}+r_{\textrm{K}}^{2}\right)^{-3/4}.\label{nier_Kerr_gor_a>0}
\end{eqnarray}
It is easy to show that the left hand side of this inequality is non-negative, since   (as we show below)
 for $r_\mathrm{K} \ge r_\mathrm{ISCO}$,
\begin{equation}
r_{\textrm{K}}\left(r_{\textrm{K}}^{2}-2a\sqrt{r_{\textrm{K}}}+a^{2}\right)\geq0.
\label{nier_Kerr_gor_a>0_nier}
\end{equation}
Indeed,  notice that  $r_{\textrm{K}}\geq\sqrt{r_{\textrm{K}}}$;
this is so because  $r_{\textrm{ISCO}}\geq1$.
But that implies $\left(r_{\textrm{K}}^{2}-2a\sqrt{r_{\textrm{K}}}+a^{2}\right) \ge \left( r_{\textrm{K}} - a \right)^2$, which is obviously non-negative. That in turn implies the non-negativity of  (\ref{nier_Kerr_gor_a>0_nier}).
Therefore the sign of the inequality (\ref{nier_Kerr_gor_a>0}) does not reverse when we calculate  the quartic power of both sides,
\begin{eqnarray}
0 & \leq & \frac{r_{\textrm{K}}^{3}}{\left(r_{\textrm{K}}\left(r_{\textrm{K}}^{2}+a^{2}\right)+2a^{2}\right)^{4}\left(r_{\textrm{K}}^{3/2}+a\right)^{4}}\times \nonumber \\
&&\left[\left(r_{\textrm{K}}\left(r_{\textrm{K}}^{2}+a^{2}\right)+2a^{2}\right)\left(r_{\textrm{K}}^{3/2}+a\right)^{4}-\right. \nonumber \\
&& \left. -r_{\textrm{K}}\left(r_{\textrm{K}}^{2}-2ar_{\textrm{K}}^{1/2}+a^{2}\right)^{4}\right].
\label{wzor1}
\end{eqnarray}

The  denominator of (\ref{wzor1}) is positive  for  $r_{\textrm{K}}\geq r_{\textrm{ISCO}}\geq1$. One can find out, using the REDUCE function of Mathematica \cite{Wolfram}\pmin{,} that the numerator  of (\ref{wzor1}) is also  non-negative\pmin{,} if $r_\mathrm{K}$ is not smaller than 1.

\paragraph{The counter-rotating case $a<0$.}
Changing   $a\rightarrow-|a|$ in the inequality  $(\ref{nier_Kerr_gor_wstawione})$,
one arrives at
\begin{eqnarray}
\lefteqn{\frac{1}{r_{\textrm{K}}^{3/2}-|a|}-\frac{2|a|}{r_{\textrm{K}}\left(r_{\textrm{K}}^{2}+a^{2}\right)+2a^{2}}\leq} \nonumber \\
&& \leq\left(a^{2}+\frac{2a^{2}}{r_{\textrm{K}}}+r_{\textrm{K}}^{2}\right)^{-3/4}.\label{nier_Kerr_gor_a<0}
\end{eqnarray}
It is easy to show that the left hand side of this inequality is nonpositive, if and only if 
 \begin{equation}
r_{\textrm{K}}^{3}-2ar_{\textrm{K}}^{3/2}+a^{2}\left(4+r_{\textrm{K}}\right)\geq0.\label{nier_Kerr_gor_a<0_nier}
\end{equation}
However for $r_{\textrm{K}}\geq 0$:
\begin{eqnarray*}
\lefteqn{r_{\textrm{K}}^{3}-2|a|r_{\textrm{K}}^{3/2}+a^{2}\left(4+r_{\textrm{K}}\right)\geq} \\
&& \geq r_{\textrm{K}}^{3}-2|a|r_{\textrm{K}}^{3/2}+4a^{2} =\left( r_{\textrm{K}}^{3/2} -|a|\right)^2+3a^{2},
\end{eqnarray*}
which is manifestly non-negative.
Thus the  left-hand side of $(\ref{nier_Kerr_gor_a<0})$ is non-negative for $r \ge r_\mathrm{K}$. Again we can calculate  the quartic power of both sides, 
\begin{eqnarray}
0 & \leq & \frac{1}{\left(r_{\textrm{K}}\left(r_{\textrm{K}}^{2}+|a|^{2}\right)+2|a|^{2}\right)^{4}\left(r_{\textrm{K}}^{3/2}-|a|\right)^{4}} \times \nonumber\\
&&\left[r_{\textrm{K}}^{3}\left(r_{\textrm{K}}\left(r_{\textrm{K}}^{2}+|a|^{2}\right)+ \right. \right. \nonumber \\
&&\left. +2|a|^{2}\right)\left(r_{\textrm{K}}^{3/2}-|a|\right)^{4}- \nonumber \\
&&\left. -\left(r_{\textrm{K}}^{3}-2|a|r_{\textrm{K}}^{3/2}+|a|^{2}\left(4+r_{\textrm{K}}\right)\right)^{4}\right].
\label{wzor2}
\end{eqnarray}
 The denominator of (\ref{wzor2}) is positive for $r_{\textrm{K}} \geq 1$. One can find out, using the REDUCE function of Mathematica \cite{Wolfram}, that the numerator is also non-negative, if $r_\mathrm{K}$ is not smaller than 6. In conclusion, in the interval of interest for counter-rotation ($r_{\textrm{K}} > r_\mathrm{ISCO} \ge 6$)   inequality $(\ref{nier_Kerr_gor_a<0})$ holds true.

\begin{acknowledgments}
This research was carried out with the supercomputer ``Deszno''
purchased thanks to the financial support of the European Regional
Development Fund in the framework of the Polish Innovation Economy
Operational Program (Contract no.\ POIG.\ 02.01.00-12-023/08).  
\end{acknowledgments}

\end{document}